\documentclass[%
aip,amsmath,amssymb, reprint
]{revtex4-1}

\usepackage{graphicx}% Include figure files
\usepackage{dcolumn}% Align table columns on decimal point
\usepackage{bm}% bold math
\usepackage[ruled,vlined]{algorithm2e}
\usepackage[utf8]{inputenc}
\usepackage[T1]{fontenc}
\usepackage{mathptmx}
\usepackage{listings}
\usepackage{tikz}
\usetikzlibrary{arrows}
\usetikzlibrary{trees}
\usepackage{midfloat}
\usetikzlibrary{decorations.pathmorphing}
\usetikzlibrary{decorations.markings}
\usetikzlibrary{automata,positioning}
\usepackage{float}
\usepackage{caption}
\usepackage{xcolor}
\definecolor{codegreen}{rgb}{0,0.6,0}
\definecolor{codegray}{rgb}{0.5,0.5,0.5}
\definecolor{codepurple}{rgb}{0.58,0,0.82}
\definecolor{backcolour}{rgb}{0.95,0.95,0.92}
\lstdefinestyle{mystyle}{
    backgroundcolor=\color{backcolour},   
    commentstyle=\color{codegreen},
    keywordstyle=\color{magenta},
    numberstyle=\tiny\color{codegray},
    stringstyle=\color{codepurple},
    basicstyle=\ttfamily\footnotesize,
    breakatwhitespace=false,         
    breaklines=true,                 
    captionpos=b,                    
    keepspaces=true,                 
    numbers=left,                    
    numbersep=5pt,                  
    showspaces=false,                
    showstringspaces=false,
    showtabs=false,                  
    tabsize=2
}

\begin{document}

\preprint{AIP/123-QED}

\title{Accelerating Coupled Cluster Calculations with Nonlinear Dynamics and Shallow Machine Learning}
% Force line breaks with \\

\author{Valay Agarawal}
\affiliation{ Department of Chemistry, Indian Institute of Technology Bombay}
\author{Samrendra Roy}
\affiliation{ Department of Energy Science and Engineering, Indian Institute of Technology Bombay}
\author{Anish Chakraborty}%
\affiliation{ Department of Chemistry, Indian Institute of Technology Bombay}
\author{Rahul Maitra}
\affiliation{ Department of Chemistry, Indian Institute of Technology Bombay}
\email{rmaitra@chem.iitb.ac.in}
\date{\today}% It is always \today, today,
             %  but any date may be explicitly specified

\begin{abstract}
The dynamics associated with the time series of the 
iteration scheme of coupled cluster theory has been
analysed. The phase space analysis indicates the presence 
of a few significant cluster amplitudes, mostly involving
valence excitations, which dictate the dynamics, while all
other amplitudes are enslaved. Starting with a few initial 
iterations to establish the inter-relationship among the
cluster amplitudes, a supervised Machine Learning scheme
with polynomial Kernel Ridge Regression model has been
employed to express each of the enslaved variables 
uniquely in terms of the master amplitudes. 
The subsequent coupled cluster iterations are 
restricted to a reduced dimension
only to determine those significant excitations, and the
enslaved variables are determined through the already
established functional 
mapping. We will show that our scheme leads to tremendous
reduction in computational time without sacrificing the
accuracy. 
\end{abstract}
\maketitle
\section{Introduction}
Coupled Cluster (CC)\cite{cc3,cc4,cc5,bartlett2007coupled}
has established itself as an accurate tool for computing 
the structure and properties of atomic and molecular
systems. In the CC method, one introduces an exponential wave
operator $\Omega$ which \textit{folds in} the effects of
excited determinants on to the reference function, often 
taken to be the Hartree-Fock determinant; $\Omega=e^T$, 
where $T$ is a sum of many-body hole-particle 
excitation operators. The unknown cluster amplitudes are
determined by projecting the similarity transformed 
Hamiltonian $G=e^{-T}He^T$ against the excited determinants.
The correlated ground state energy is computed by evaluating
the expectation value of effective Hamiltonian with 
respect to the chosen reference function,
$E_{corr} = \langle H_{eff} \rangle = \langle e^{-T}He^T \rangle $.
Due to the exponential wave operator, the amplitude
determining equations are nonlinear and hence, 
one almost universally employs the iterative scheme to 
find the fixed points. The amplitude $t_\mu$ associated
with the excitation operators, $T_\mu$ is determined by
demanding $g_\mu=0$. Here $g$ represents the amplitudes of
the similarity transformed Hamiltonian $G$, and $\mu$ is the
combined hole-particle labels associated with the excited
function. Clearly, in a CC scheme with single and double
excitations (CCSD), the dimension of $t$'s becomes 
($n_on_v+n_o^2n_v^2$), where $n_o$ and $n_v$ are 
the number of hole and particle orbitals respectively, 
and determining these many amplitudes 
requires an iterative $n_o^2n_v^4$ scaling. 

In a recent paper\cite{agarawal2020stability} by the present
authors, a \textit{posteriori} analysis of the time series
associated with the iterative scheme of a different version
of the CC theory was presented. The authors introduced 
an input perturbation to the amplitude determining equations
to probe the nonlinearity associated with the discrete-time 
dynamics of the iterative scheme. It was established that such 
equations show interesting features of the universal 
chaotic dynamics characterized by a full period-doubling bifurcation cascade, and that there exist nontrivial 
inter-relationships among the different cluster amplitudes.
As such, the macroscopic features of the 
dynamics is solely dictated by a few large significant
cluster operators \textit{(vide infra)}, which span a 
much smaller space. These cluster amplitudes behave as 
the order parameters, while the remaining cluster 
amplitudes are enslaved under the former set. 
Taking insight from 
nonlinear dynamics and Synergetics\cite{Haken_1989, haken1982slaving, Haken1983}, the authors predicted a 
mapping of the enslaved amplitudes in terms of those 
significant \textit{master} amplitudes. 
In this work, we further extend the analysis and show that
it is indeed possible to exploit the master-slave
multivariate dynamics to numerically map the enslaved
amplitudes as functions of those significant ones. 
In order to establish such a mapping, we have employed a
supervised machine learning (ML) strategy, based on the
Kernel Ridge Regression (KRR)\cite{murphy2012machine} model
and come up with a hybrid CC-ML algorithm for solving CC
 theory with excellent savings of computation time.

One may also note that the ML algorithm is employed 
to establish the synergy and inter-relationship among 
the cluster operators, which are predicted by the
multivariate 
nonlinear dynamics. The methodology is solely based on the
time-series dynamics associated with the iteration process,
and is specific to the individual many-body system. Hence,
unlike other ML methods, particularly those based on the neural network which provides data-driven solutions
to CC theory,
\cite{miller2019regression, margraf2018making,townsend2019data, peyton2020machine,folmsbee2020assessing, schran2019automated, welborn2018transferability, schutt2019unifying}, our scheme
does not require \textit{any} prior computation to train 
the model. Contrary to that, our method is physically 
motivated and its origin is grafted in the dynamics 
of the iteration process, which allows one to map the
the dependency of the cluster operators via supervised ML. 

The paper is organized as follows: In Sec II we establish,
via a phase space analysis, that a few large cluster amplitudes dictate the dynamics, and that the variation of all other amplitudes is suppressed. This master-slave
the relationship among the cluster operators allow us to express
the dependent enslaved variables as functions of the master
amplitudes and we shall outline the hybrid CC-ML iterative
scheme in Sec. III. We present the essential aspects of the
KRR ML model in III.1, and the following section, 
we show how one may construct the CC equations for selected variables 
with much lower scaling. We will present the efficacy of our scheme 
in Sec. IV, and conclude our findings in Sec V. 

\section{Establishing Master-Slave Dynamics} 
\begin{figure}[!h]
\includegraphics[width=\linewidth]{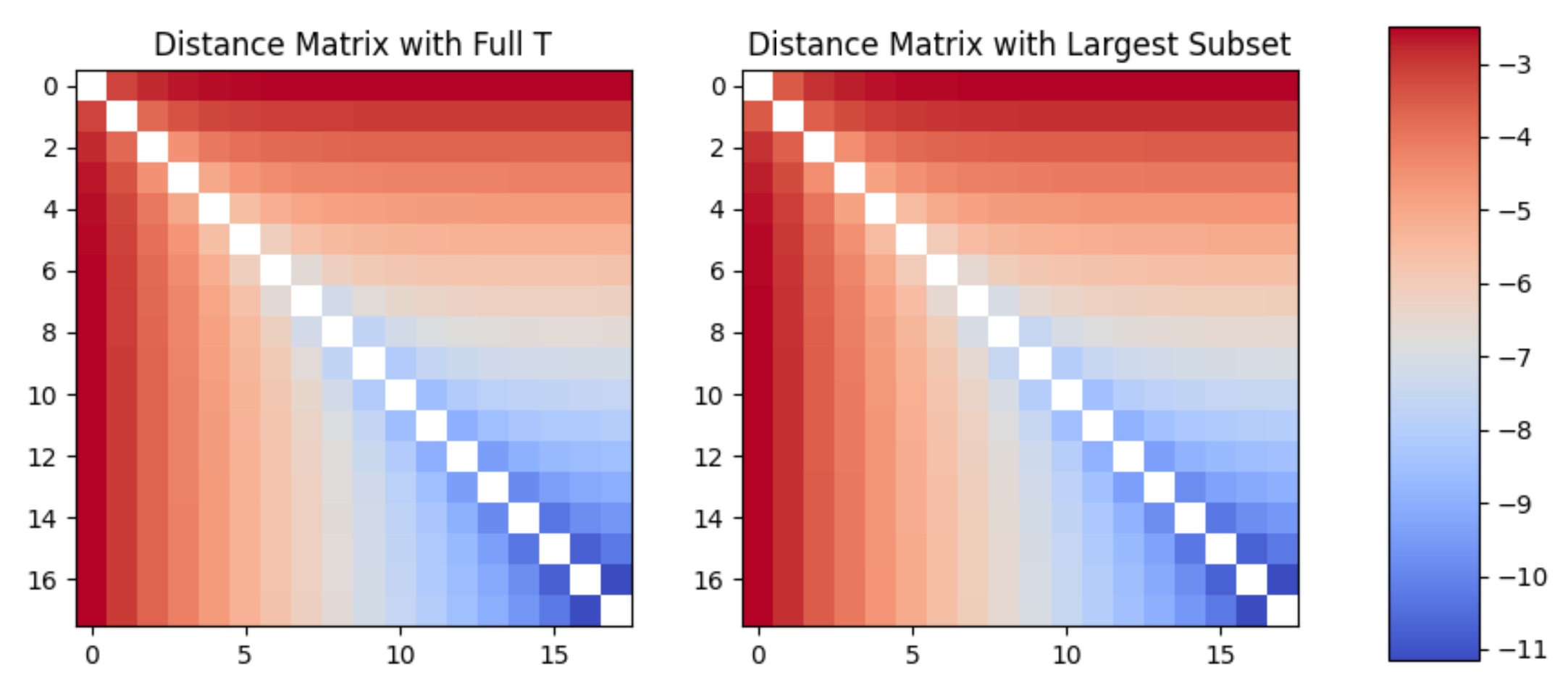}
\caption{Distance matrix (in logarithm scale) associated with the iteration time series. The bias in the distance matrix towards the right lower corner signifies the drift towards the fixed point. Note that the diagonal will be zero, so there is no graph plotted there. Both the vertical and horizontal axes represent iteration steps.}
\label{fig:distance-matrix}
\end{figure}

In order to establish the significance of the master 
amplitudes in the dynamics, we analyze the 
phase space trajectory associated with the
iteration scheme. The iteration procedure is considered
as a time-discrete dynamics. To visualize the
multidimensional phase space in two dimensions, 
one may resort to the distance matrix (DM)\cite{eckmann1987recurrenceplot,marwan2007recurrence,RecurrencePlot}, which is 
defined as $DM_{i,j}=||\vec{x_i}-\vec{x_j}||$. 
Here each iteration is embedded as a single time step 
and $\vec{x_i}$ and $\vec{x_j}$ 
are the amplitude vectors at time step $i$ and $j$ respectively. 
Thus the $\vec{x_k} = (t_{1_k}\oplus t_{2_k})^T$. DM represents 
the closeness of the trajectory at two different time steps.
Figure \ref{fig:distance-matrix}A depicts the DM for $H_2O$ at
bond length = 2.6741 Bohr, bond angle = 96.774$^\circ$ in 
cc-pVTZ basis. The bias 
in the DM towards the lower right corner indicates that the
dynamics associated with the iterative scheme is convergent 
towards a unique set of fixed points. Such DM, however, 
under large input perturbation shows repetitive or chaotic phase space trajectory, as previously established by 
Agarawal et al.\cite{agarawal2020stability}. \\

To establish our hypothesis of a slaving dynamics, we have 
further constructed the amplitude vectors with only those 
cluster amplitudes whose magnitudes are greater than a 
predefined large threshold $\epsilon$. We will generically
denote these amplitudes as $t^L$, where the superscript
denotes that these set of amplitudes include only large 
ones and would refer to this set 
as the \textit{largest subset} (LS). Note that the LS
may include both one- and two-body excitations depending
on the multi-reference nature of the molecular system under
consideration. For CC theory with singles, doubles, and
triples excitation scheme (CCSDT), the LS may include triple
excitation as well. From a theoretical perturbation
point of view, it is obvious that the
excitations belonging to the LS mostly involve valence
electrons. In this analysis, the choice of the 
threshold $\epsilon$ for selecting the LS is taken
arbitrarily at 0.02 at the moment,
although a more robust choice may be made through
maximization of mutual information. One may 
note that $\{t^L\}$ is only a very tiny subset of the
full set of cluster operators, while 
all other cluster amplitudes with smaller magnitudes are 
grouped into the smaller subset (SS), denoted by $\{t^S\}$. 
%\begin{figure}[!h]
%\includegraphics[width=\columnwidth]{change_in_t.PNG}
%\caption{Difference between successive iterations}
%\label{fig:t-distance}
%\end{figure}.
In what follows, we further construct 
the DM by taking into account only the LS amplitudes:
$DM_{i,j}=||\vec{x^L_i}-\vec{x^L_j}||$, where 
$\vec{x^L_k} = (t_{1_k}^L\oplus t_{2_k}^L)^T$ and the corresponding DM is shown in Figure \ref{fig:distance-matrix}B.
It is evident from the two nearly identical DM's that the 
LS qualitatively and quantitatively replicate the dynamics exhibited by the full set of cluster amplitudes. A lower
choice of the threshold $\epsilon$ makes the DMs 
exactly \textit{identical}. The choice of the molecule and 
the threshold discussed above is presented as a prototypical
case. A similar analysis can be made for other systems as 
well. In other words, such an analysis is not system-specific. In a previous paper\cite{agarawal2020stability}, 
a similar analysis was put forward for studying
the iteration dynamics under an 
input perturbation for regions where linear stability
is lost. Thus the current work may be considered as a 
generalization for multivariate iteration dynamics for 
regions which are characterized by a unique set of fixed-point solutions.\\

\section{Synergetic Mapping of the Largest Subset to the Smaller Subset and the Feedback Coupling:}
It has been established in the previous section via the full and the reduced space DMs that the macroscopic 
features of the
iteration process is almost quantitatively
governed by the LS. As such, the variation of very
large number of smaller amplitudes are suppressed and thus their
microscopic sub-dynamics is asymptotically negligible. 
The amplitudes belonging to the LS, $\{t^L\}$, 
may be considered as the order parameter of the system, which
enslave the smaller ones. Let us denote the dimension 
of the LS by $n_l$. In such multivariate cases, in the 
regions away from fixed-point equilibrium, one may map 
the smaller amplitudes as unique functions of the 
order parameters. In other words, one may consider 
the elements of the LS, $\{t^L\}$, as independent 
variables of the dynamics, while those belonging 
to $\{t^S\}$ may be considered
as the dependent slave variables. The dimension of the 
enslaved smaller amplitudes is denoted by $n_s$. Note 
that $n_l \textless \textless n_s$. For convergent 
regions with fixed-point equilibrium, we conjecture 
that for a given time step, one may write:
\begin{equation}
t^S_{\mu, k}=F_\mu(\{t^L_k\})
\label{eq1}
\end{equation}
where $k$ represents a particular iteration step 
and $\mu$ is the shorthand notation of hole and 
particle indices associated with the excitation.
Eq.\ref{eq1} is in principle exact for away from 
fixed point equilibrium regions and can be 
determined analytically for systems with few variables. 
However, it is impossible to predict this exact mapping 
for high dimensional cases. We have employed a supervised ML 
model to construct the functional 
form of $F_\mu$. In what follows, we have performed a few 
initial CC iterations in full dimension (space spanned by
$\{t^L\oplus t^S\}$) and considered the cluster amplitudes
of these initial iterations as the training data set. 
The polynomial regression based supervised ML
model has been employed to 
express the small dependent amplitudes as functions of the
independent order parameters. We have assumed that the exact
inter-relationship among these two sets of amplitudes are 
established within the first few iterations and it remains 
unchanged
over the subsequent iteration steps. Our numerical results
would strongly suggest this to be a good approximation. 
Only the cluster amplitudes belonging to the LS, 
on the other hand, are determined via the conventional 
algebraic or diagrammatic techniques, where both sets of
cluster operators couple. The coupling of the cluster
operators belonging to the
smaller subset (which are determined via the ML) to
the equations of the amplitudes of LS may be termed
as the \textit{feedback coupling}. We shall show that the
construction of the diagrams for the cluster amplitudes 
belonging to the LS can be achieved at a much 
cheaper scaling than the conventional way. This 
reduction of the independent degrees of freedom leads 
to the enormous savings in computational time. 

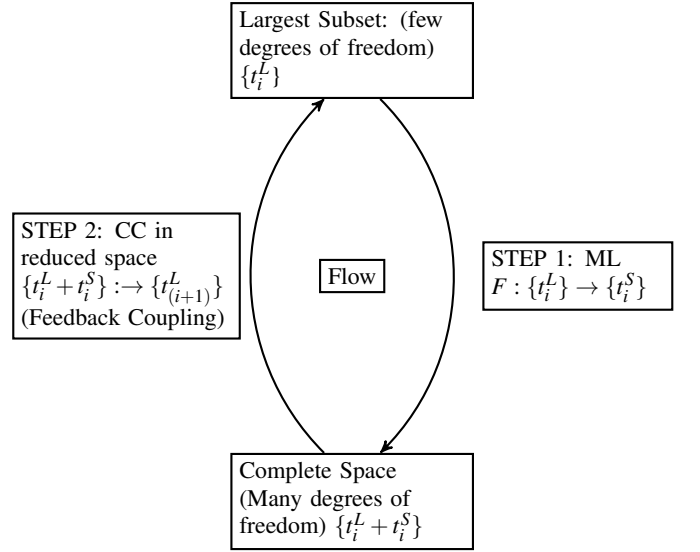
\begin{figure}
    \centering
\begin{tikzpicture}
  [
    ->,
    >=stealth',on grid,
    auto,node distance=3cm,
    thick,
    main node/.style={rectangle, draw}
    ]
    \node[main node]      (maintopic)                              {Flow};
\node[main node]        (LS)       [above of= maintopic,text width=3cm] {Largest Subset: (few \newline degrees of freedom) \newline $\{t^L_i$\}};
\node[main node] (ML) [right of= maintopic,text width=2.3cm] {STEP 1: ML $F: \{t^L_i\}\rightarrow \{t^S_i\}$};
\node[main node] (FT) [below of= maintopic,text width=3cm] {Complete Space (Many degrees of freedom) $\{t^L_i+t^S_i\}$};
\node[main node] (CCT) [left of= maintopic,text width=2.8cm] {STEP 2: CC in \newline reduced space\newline  $\{t^L_i+t^S_i\} :\rightarrow \{t^L_{(i+1)}\}$  (Feedback Coupling)};
\draw[->] (FT.120) to[out=135, in=-135] (LS.-120);
\draw[->] (LS.-60) to[out=-45, in=45] (FT.60);
\end{tikzpicture}
    \caption{The circular causality loop employed in this work. The subscript quantity denotes the iteration time step. The numerous enslaved variables are determined via the functional mapping (step 1) with polynomial regression that has been established \textit{apriori} via ML. The CC iterations are restricted only to determine the elements belonging to the LS (step 2).}
    \label{fig:my_label}
\end{figure}

In short, the overall scheme 
starts with $m$ conventional iterations, which are used to
train the model to determine the function $F$. The 
accuracy of the model depends on the value of $m$. This is
followed by the CC-ML hybrid iterative algorithm, which is
schematically represented in Figure \ref{fig:my_label}. 
The hybrid CC-ML algorithm involves two major steps. 
Step 1 performs the forward mapping of the LS 
amplitudes (order parameter) 
on to the smaller enslaved subset via Machine learning (ML),
and in step 2, the enslaved variables provide the feedback
coupling to determine the order parameters (LS). This 
is commonly referred to as the \textit{circular causality}
in Synergetics. In the following section, we briefly present
the idea of Kernel Ridge polynomial Regression (KRR) method
which maps the order 
parameters to the enslaved variables (step 1), and refer 
the readers to other ML
texts\cite{murphy2012machine} for details. The next 
subsection will deal with the essential aspects of CC 
diagrammatic construction for the LS of   
amplitudes (step 2).

\subsubsection{Brief Account of Kernel Ridge Regression based ML Model:}

As previously mentioned, the forward mapping (step 1)
of the amplitudes of the LS to the smaller subset is done
via polynomial Regression-based supervised ML
coupled with Ridge Kernelization. This 
regression algorithm is based on the linear regression model, 
where the nonlinear terms of the independent variables are 
also considered as independent parameters. As we discussed,
the initial few iterations were performed in the full
space spanned by the entire $\{t^L\oplus t^S\}$ 
amplitudes and these exact amplitudes were used for 
training the model. As expected for any supervised machine 
learning algorithm, the accuracy of the model would
depend on the dimension of the training data set ($N_T$).
The linear regressor fits a set of independent variables to
a set of dependent variables using the inter-dependency 
of the data structure. Such a generalised regression
relationship can be written in a matrix form as 
$T^S = T^L \beta$.  
Here $T^L$ is the matrix of independent variables, which
are the cluster amplitudes belonging to the LS.
%\textcolor{red}{Should we mention the size of matrix in eq 2,3,4}
\begin{equation}
T^L = 
  \begin{pmatrix}
    1 & t^L_{11} & t^L_{21} & ... & t^L_{{n_l} 1}\\
    1 & t^L_{12} & t^L_{22} & ... & t^L_{{n_l} 2}\\
    1 & t^L_{13} & t^L_{23} & ... & t^L_{{n_l} 3}\\
    ... & ... & ... & ... & ...\\
    1 & t^L_{1m} & t^L_{2m} & ... & t^L_{{n_l} m}\\
\end{pmatrix}  %\;\;\;\;\;\;\;\;m *(\nu+1)
\end{equation}
Here each row signifies the independent cluster 
amplitudes for a fixed iteration. With 
$m$ number of training iterations were performed to 
construct the $T^L$ matrix, it is of the dimension
$m*(n_l+1)$. The extra column in $T^L$ is added to take 
care of the intercept term.
%where $m$ is the number of training sets, $\nu$ is the size of the independent variable vector

Similarly, the coefficient matrix $\beta$ may be defined
as:
\begin{equation}
\beta = 
  \begin{pmatrix}
    \beta_{01} & \beta_{02} & \beta_{03} & ... & \beta_{0 {n_s}}\\
    \beta_{11} & \beta_{12} & \beta_{13} & ... & \beta_{1 {n_s}}\\
    \beta_{21} & \beta_{22} & \beta_{23} & ... & \beta_{2 {n_s}}\\
    ... & ... & ... & ... & ...\\
    \beta_{{n_l} 1} & \beta_{{n_l} 1} & \beta_{{n_l} 2} & ... & \beta_{{n_l} {n_s}}\\ 
\end{pmatrix} % \;\;\;\;\;\;\;\; (n_l+1)*n_s
\end{equation}
where $n_S$ is the size of the dependent variable vector
$T^S$, which is given by:
\begin{equation}T^S = 
     \begin{pmatrix}
    t^S_{11} & t^S_{21} & ... & t^S_{n_s 1}\\
    t^S_{12} & t^S_{22} & ... & t^S_{n_s 2}\\
    t^S_{13} & t^S_{23} & ... & t^S_{n_s 3}\\
    ... & ... & ... & ...\\
     t^S_{1m} & t^S_{2m} & ... & t^S_{n_s m}\\
\end{pmatrix}  % \;\;\;\;\;\; m*n_s
\end{equation}
Starting with a guess coefficient matrix is $\hat{\beta}$,
the coefficients may be obtained by minimizing the loss function $\eta^T \eta$, where the error $\eta$ is given by:
$\eta = T^S - \hat{T^S}$
where $\hat{T^S}=T^L\cdot \hat{\beta}$. Here $\hat{T^S}$ is
the predicted small subset amplitudes. 

Often, a linear model may not be enough to capture the 
features of a data set, and hence, a quadratic, cubic or
some other polynomial model is used. Naturally, 
it leads to 
increase the dimension on the independent variables, and
hence provide better flexibility to have a better chance 
of getting a good fit. One may increase the total feature
vectors from $n_l$ to much higher dimensional space by including
polynomial terms, and then each of the linear and nonlinear
terms are treated as independent variables 
to solve using the linear regression
technique. One may define a function
$\phi$, which maps the given set of feature vectors to a 
higher dimension space by including the non linear terms: 
\begin{equation}
    \phi: \{t^L_\mu, \mu \in n_l\} \rightarrow\{t^L_\mu,t^L_\mu t^L_\nu,...,(t^L_\mu)^d, \mu,\nu\in n_l\}
\end{equation}
where $d$ is the degree of the polynomial. One may minimize
the loss function with respect to $\hat{\beta}$ to arrive 
at an expression 
$\hat{\beta_\mu}=\phi^T (\phi \phi^T)^{-1} (T^S_\mu)$.
However, a direct expansion to a nonlinear expressions and 
determination of the coefficient matrix by evaluating the
above expression is computationally expensive, and hence,
one often uses Kernelization technique instead, which 
allows evaluating the expression without explicit 
knowledge of the function $\phi$.

According to the Mercer theorem, one may
define the Kernel Function $K=\phi \phi^T$
for every symmetric positive definite matrix.
Thus one may write $\hat{\beta_\mu}=\phi^T (K)^{-1} (T^S_\mu)$. Once the model is trained, the SS 
amplitudes for the $i$-th iteration are
predicted using the coefficient matrix and the LS vector obtained from the same iteration.
\begin{equation}
    t_{\mu,i}^S (predicted) = \phi(\{t^L_i\}) \hat{\beta}_\mu
\end{equation}

Towards this, let us define a new kernel function 
$K_L$, which takes all
the previous training amplitudes and the new set of LS
amplitudes of a given $i-th$ iteration to predict the
new set of amplitudes $t^S$. Thus, 
$K_L = \phi \phi(t^L_i)^T$, 
where the function $\phi$ appearing to the right is a 
function of the new LS amplitudes, which has been
shown explicitly. 
Using $K_L$, one may obtain the predicted elements 
of the small component as: 
\begin{equation}
    t^S_{pred} = (K_L)^TK^{-1}T^S
    \label{eq:KernelReg}
\end{equation}
Here the subscript indicates that these are the 
predicted SS amplitudes. 

Sometimes, the model may get unphysical
underfitting or overfitting due to 
erroneous weight in the 
training data set. In order to control that,
a regularization parameter is often introduced 
which penalizes the model each time a certain
term gets unphysical weight. 
Thus one may modify Eq \ref{eq:KernelReg} by 
adding a regularization term to \begin{equation}
     t^S_{pred} = (K_L)^T(K+\lambda I)^{-1}T^S
     \label{eq:ridge}
\end{equation}. 
Following the conventional notation used in
Ridge regularization, we will denote the 
regularization parameter with $\alpha$. Here
$\lambda=\alpha/2$. The value of 
$\lambda(\alpha)$ manages overfitting at the
cost of rate of learning. Very small value
of $\alpha$ trains the model too fast and 
overfits the data points, which results in 
slight inaccuracy with few training data 
sets. On the other hand, a large $\alpha$ 
slows down the learning process, and the 
model takes a larger number of training data 
sets to produce accurate results. Thus an
optimized value of $\alpha$ is warranted 
for the numerical accuracy of the model.

In Eq. \ref{eq:ridge}, we note that the 
quantity $(K+I \lambda)^{-1}T^S$ can be computed 
only once for all after the training data set
is produced. Thus all the $t^S$ amplitudes 
may be computed each iteration via a
single matrix multiplication of $(K_L)^T$ and $(K+I \lambda)^{-1}T^S$, which results in huge
computational time savings.

It should be noted that the method is now readily 
available in ML libraries and can directly be 
used\cite{scikit-learn}. Therefore, while training, the 
inputs are a matrix of independent variables ($\{t^L\}$ 
amplitudes), and a matrix of the dependent variables
($\{t^S\}$ amplitudes). While predicting, the input is 
simply the new LS cluster amplitudes, and the output is 
the vector of dependent SS amplitudes.

\subsubsection{Exact Determination of Order Parameters via Coupled Cluster Theory:}
As previously mentioned, the LS amplitudes are determined
using CC theory (step 2). Once the SS of amplitudes is 
generated, one may update the LS cluster amplitudes via 
usual Jacobi or conjugate gradient methodology. 
\begin{equation}
    t^L_{\mu, i+1}=t^L_{\mu,i}+\frac{g^L_{\mu,i}}{\Delta D_\mu} \;\;\;\;\;\;\;\; \forall \mu \in LS
\end{equation}
\begin{figure}[!h]
    \centering
    \includegraphics[width=\linewidth]{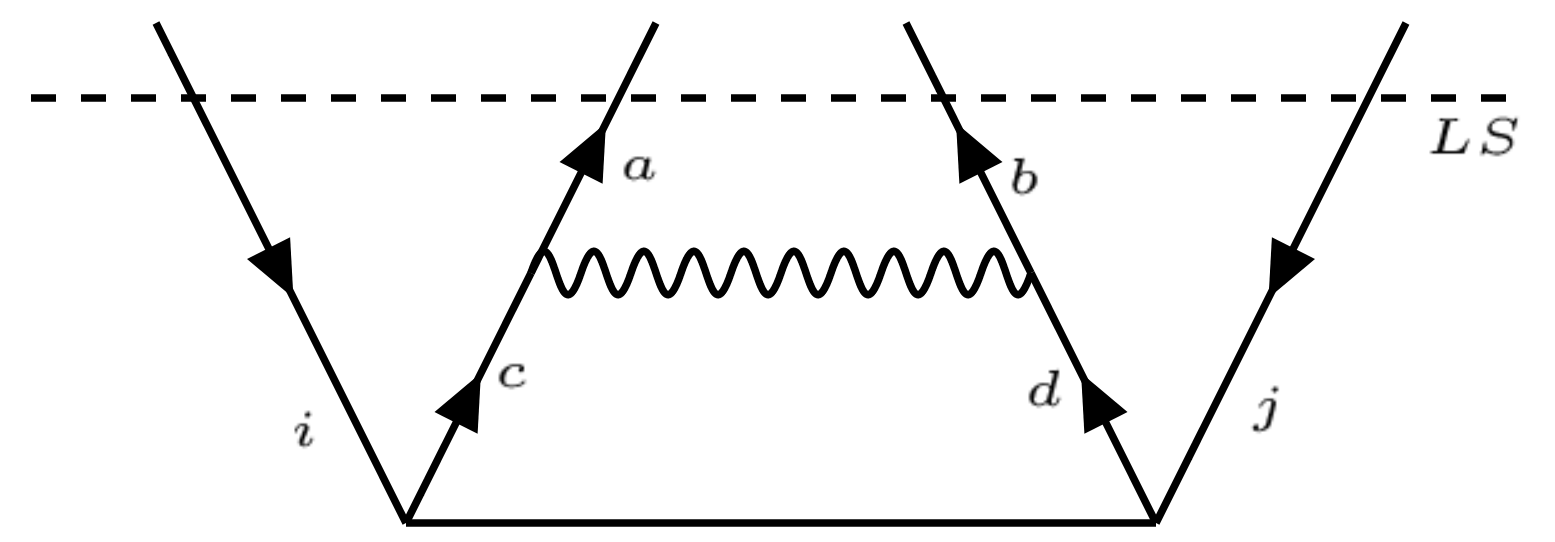}
    \caption{A representative CC linear diagram. The wiggly line denotes the two electron interaction and the solid horizontal line denotes a $T_2$ operator. The set of external uncontracted indices is restricted only to those sets which constitute the LS and is depicted by the dashed line, enabling a scaling reduction to $n_ln_v^2$ from the usual scaling of $n_o^2n_v^4$.}
    \label{fig:linear}
\end{figure}
\begin{figure}[!h]
    \centering
    \includegraphics[width=\linewidth]{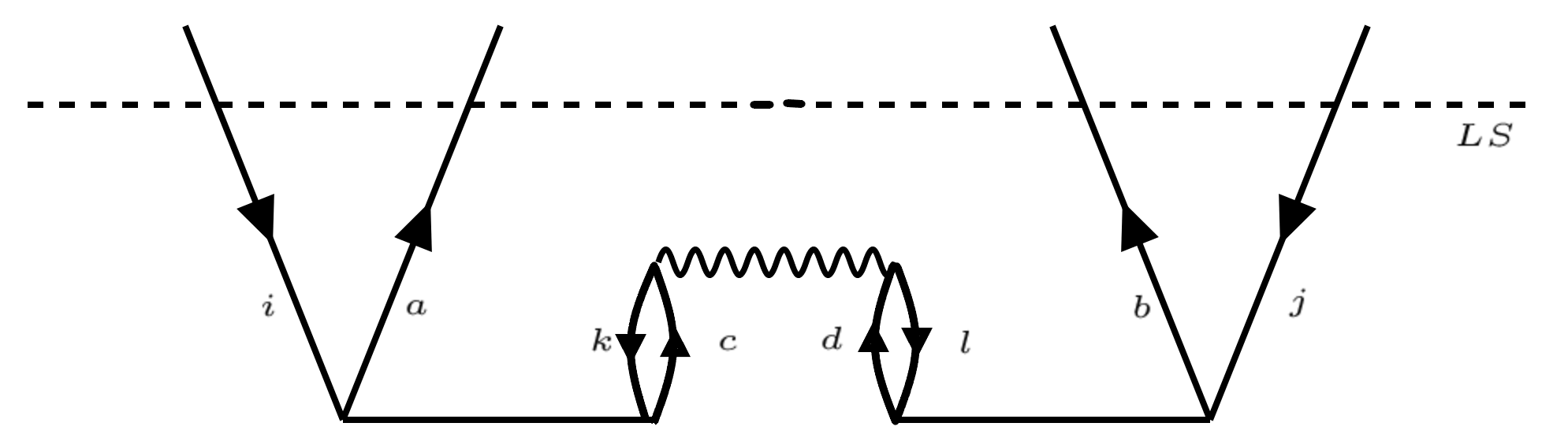}
    \caption{A representative CC nonlinear diagram. The set of uncontracted indices are restricted to only those which belong to the LS, enabling construction of the entire diagram at a scaling of $n_l n_o^2 n_v^2$ without constructing it in a step-wise manner.}
    \label{fig:non-lin}
\end{figure}
\begin{figure*}
    \begin{minipage}{\textwidth}
    \centering
    \includegraphics[width=\linewidth]{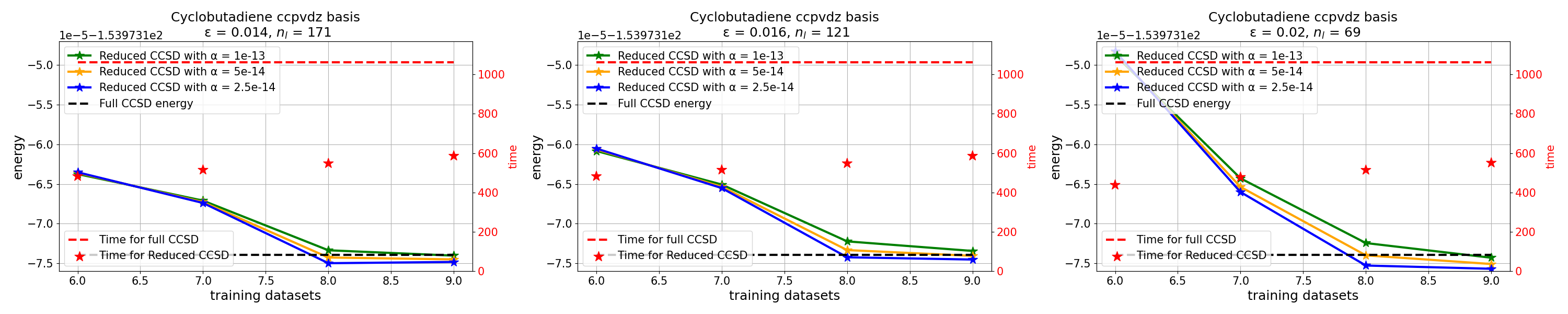}
    \captionof{figure}{Predicted energies and the time taken for cyclobutadiene in cc-pVDZ basis (C-C bond = 1.88 Bohr, C-H bond = 1.313 Bohr) as functions of the training data set size (number of training iterations, $m$). The LS dimension, $n_l$, is explicitly mentioned for different values of $\epsilon$, and there are total 112720 \textit{nonzero} cluster amplitudes. With sufficiently small $\alpha$, one gets sub $\mu E_h$ accuracy with 45$\%$-50$\%$ reduction in overall computation time. Note that the grid size is 5 $\mu E_h$. Change of $\alpha$ does not require any extra time for the calculations. }
    \label{fig:cyclobutadiene}
    \end{minipage}
\end{figure*}
Here the orbital labels '$\mu$' associated with
the excitations are necessarily restricted to
those belonging to the LS and hence it carries 
the superscript $L$. The quantity $g^L_\mu$ is
the residue associated with the excitation 
$\mu \in LS$, and $\Delta D_\mu$ is the usual
orbital energy difference. One may note that
$g^L_\mu$ can be evaluated either algebraically
or diagrammatically via the
Baker-Campbell-Hausdorff (BCH) multi-commutator 
expansions where the entire set of cluster 
amplitudes contribute. However, the set of
external (uncontracted) indices take up only
those tuples which belong to the LS. We discuss
below how such a restriction simplifies the 
construction of CC diagrams at a cheaper 
computational scaling. \\

Let us consider the most expensive CC diagram,
where two virtual orbital indices contract. In
the conventional CC theory, this diagram scales
as $n_o^2n_v^4$. In our modified scheme, we
construct such diagrams for only those 
excitations where the set of uncontracted 
indices belong to the LS. Fig. \ref{fig:linear} shows
the diagram where the dotted line 
labelled $LS$, intersecting the uncontracted 
indices indicates the set $\{ijab\}$ 
necessarily belongs to the $LS$. Note
that the dimension of such amplitudes is $n_l$.
However, the cluster operators that contribute
to this diagram, $t_{ijcd}$, can take any 
hole and particle indices. That implies
$t_{ijcd}$ can belong to either the LS or SS.
Thus, the total scaling to construct the
diagram is simply $n_l n_v^2$. Note that the
circular causality demands the coupling of
smaller amplitudes to the equations for the LS
amplitudes, and hence there is non-trivial
coupling of the different amplitudes to ensure
the clustering effect. 

We now turn our attention to a representative 
nonlinear diagram. Let us consider the diagram 
as shown in Fig. \ref{fig:non-lin}. Note that 
in such a case,
the external indices arise from two different 
cluster amplitudes and hence it is convenient
to construct the entire diagram at once. Like
the linear diagram discussed previously, the 
external indices are restricted to only those
sets of orbitals which belong to the LS. That
means the entire diagram may now be constructed
at once with a scaling of $n_l n_o^2n_v^2$ at
worst. On top of the scaling reduction, each
of the nonlinear terms can be constructed 
at a single step, without any requirement of
constructing the optimal intermediate. This 
further reduces the number of matrix operations
by almost a factor of half. One may note that
the cluster operators contributing to the
nonlinear terms may belong to either the LS
or the SS, and one may further judiciously
approximate by including only those nonlinear 
terms where all the cluster operators belong 
to the LS. That implies that one may consider
the cluster amplitudes of the SS up to the 
linear terms to provide the feedback coupling.
We shall demonstrate in the next section that 
the hybrid CC-ML algorithm would enable us to 
compute the energetics of
molecular systems with sub-microHartree ($\mu E_h$)
accuracy with tremendous savings in computational time.  
\section{Results}
\begin{figure*}
    \centering
    \includegraphics[width=\textwidth]{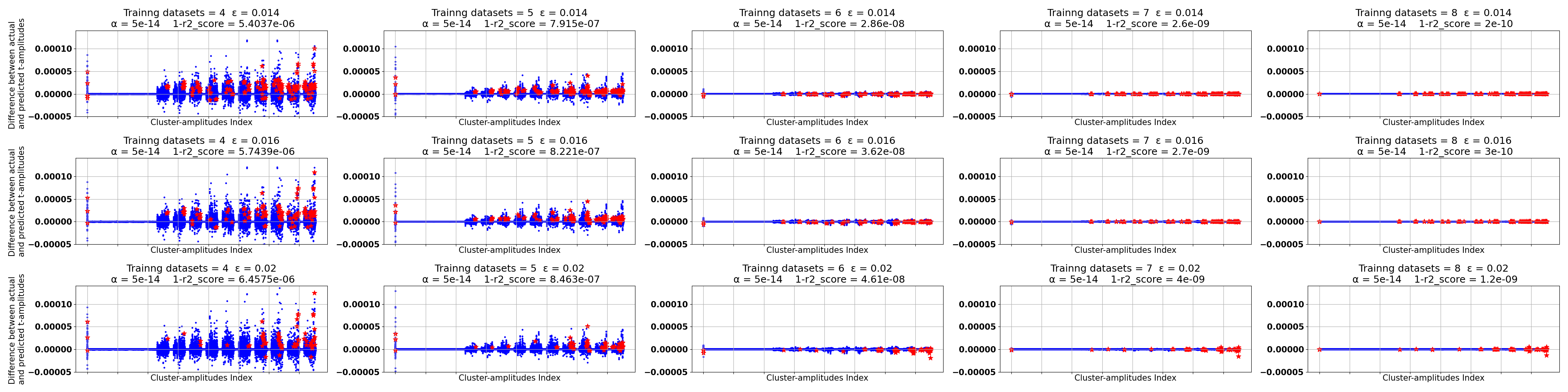}
    \caption{Difference between predicted and actual cluster amplitudes vs cluster amplitude index. Red dots denote the difference between the predicted and actual LS amplitudes, while the blue dots are for the SS amplitudes. Note the increase in accuracy of r2 score, which is represented as the difference from 1. r2 score = 1 represents perfectly accurate prediction.}
    \label{fig:scatter}
\end{figure*}
In this section, we demonstrate the efficacy 
of our scheme with a few pilot numerical examples. 
We shall show that it is indeed possible to 
achieve very high accuracy with the
hybrid CC-ML algorithm discussed previously, with a
tremendous reduction in overall computation time.
We first consider the case of cyclobutadiene molecule 
(C-C bond length = 1.88 Bohr and the C-H bond length =
1.313 Bohr), in the cc-pVDZ basis. The convergence of 
the predicted energy with respect to the training data 
set size $m$ is presented in Fig. \ref{fig:cyclobutadiene}
with three different threshold value $\epsilon$. 
Note that a lower
value of the threshold $\epsilon$ signifies a larger LS
dimension, $n_l$. Clearly, a larger LS dimension 
provides much more flexibility to fit the SS amplitudes 
as functions of those of the LS. As evident from Fig.
\ref{fig:cyclobutadiene}, it is indeed the case as the
predicted energy gets more accurate with lower $\epsilon$.
The exact CCSD energy is shown with the black dashed line.
Furthermore, for each value 
of $\epsilon$, we have plotted the predicted energy as a
function of $n_l$ for three different regularization 
parameter $\alpha$. Note that for any sufficiently small
$\alpha$, the predicted energy reaches sub $\mu E_h$
accuracy within a training data set size of eight.
However, as an artifact of any regression technique, 
too low value of alpha leads to overfitting the model 
due to fast learning process, which may lead to slight
inaccuracy and non-monotonic nature of energy convergence,
particularly with lower training set size. On the other
hand, a relatively larger $\alpha$
leads to a slow learning process, which often results 
in a large number of training steps to achieve similar 
accuracy.

We have further plotted the overall time taken by our
hybrid CC-ML algorithm for different training set
dimensions, and compared those against the
time taken for the exact CCSD calculations with same 
convergence threshold and same computer architecture. The
time scale is shown on the right-hand side of each plot,
and the time taken by exact CCSD is denoted by the red 
dashed line, while the time taken by our method is shown 
by red dots. Note that for a given $\epsilon$ and fixed 
training data set dimension $m$, the time taken
by our method for different $\alpha$ is almost 
the same and hence they are denoted by a single 
red dot in each plot. Clearly, with for all the 
threshold values $\epsilon$ reported here, 
our method reaches sub $\mu E_h$ accuracy (with training 
set dimension $m=8$) with almost 48$\%$ less time 
(with $\epsilon=0.016$, and even better time 
savings for larger $\epsilon$ values) compared 
to the time taken by the exact CCSD methods.
With $m=9$, our model takes about 45$\%$ less time
(with $\epsilon=0.016$) compared to the exact method 
for similar accuracy.

In order to gain insights into the accuracy of the
predicted cluster amplitudes via the ML algorithm, 
in Fig \ref{fig:scatter}, we have plotted the 
difference between our predicted cluster amplitudes 
and the exact converged amplitudes. Therefore, a 
plot with less scattered points suggests a good fit. 
Three different rows of the plot
are for three different $\epsilon$ values, and for each
$\epsilon$, we have plotted the difference in the
amplitudes for different training set size with 
$\alpha=5 \times 10^{-14}$.
Clearly, for any given $\epsilon$, as we increase the
training set size, the scattered points get more 
aligned along the $y=0$ line, signifying a high 
accuracy in the prediction.
As evident from Fig. \ref{fig:my_label}, the 
predicted SS amplitudes contribute to determining 
the LS amplitudes via the circular causality loop. 
Therefore, any error in prediction leads to 
accumulation of the error, which in turn contaminates 
both the LS and SS amplitudes. We note from the
scatter plot that both LS amplitudes, denoted by the red
dots and the SS amplitudes (denoted by the blue dots) get
extremely accurate with sufficient training set size. 
This results in excellent accuracy in the predicted 
energy, as we had shown earlier.
The accuracy in the prediction is also validated by
computing the r2 score, which tends to one as we 
include more training sets into the algorithm.

\begin{figure}[!h]
    \centering
    \includegraphics[width=\linewidth]{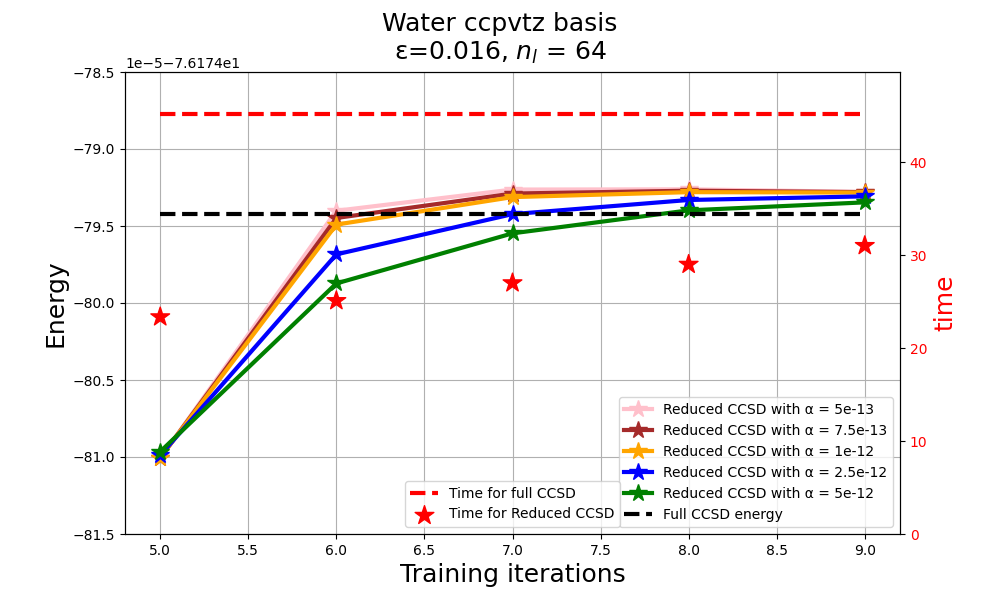}
    \caption{Predicted energies and the time taken for Water in cc-pVTZ basis (O-H bond = 2.6741 Bohr, H-O-H angle = $96.774^\circ$) as functions of the training data set size (number of training iterations, $m$). Note that the grid size is 5 $\mu E_h$.}
    \label{fig:water}
\end{figure}
\begin{figure}[!h]
    \includegraphics[width=\linewidth]{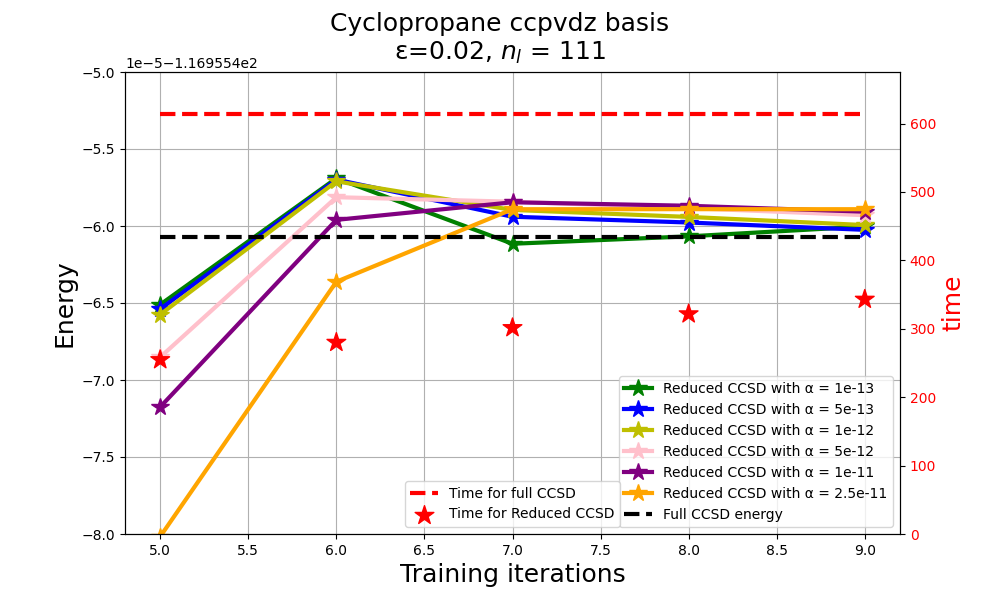}
    \caption{Predicted energies and the time taken for cyclopropane in cc-pVDZ basis (C-C bond = 2.13 Bohr, C-H bond = 1.53 Bohr, C-C-C bond angle = $60^\circ$, H-C-H bond angle = $114.6^\circ$) as functions of the training data set size (number of training iterations, $m$). Note that the grid size is 5 $\mu E_h$}
    \label{fig:cyclopropane}
\end{figure}
To prove the efficacy of the algorithm across 
different molecules, we have also shown the accuracy 
of our method for water in 
cc-pVTZ basis (Fig. \ref{fig:water}) and cyclopropane in 
cc-pVDZ basis (Fig. \ref{fig:cyclopropane}). The 
corresponding geometries are given along with the
figures. In both the cases, one gets $\mu E_h$ level
accuracy within 7 training set size and at 35$\%$-45$\%$
reduction in overall computation time. The slight
inaccuracy and non-monotonic nature of the plot with 
very small $\alpha$ for cyclopropane is an artifact 
of overfitting due to the fast learning process. 
One may note that for larger molecules, further 
savings in computation time is expected
as the LS dimension, $n_l$ grows sub-linearly with the 
system size, which makes the entire scheme a scalable one.
For systems which take fewer iterations to converge, the
mutual dependency among the cluster operators during the 
iteration process is established much earlier, which 
results in fewer training iteration for accurate results.

One may also note that the model involves no hidden 
training costs. Unlike most of the CC methods based on 
ML which pre-train their models on thousands of 
molecules to predict, our method 
\textit{does not require any prior training} on 
any molecular data set, and is entirely based on 
the physically motivated method of
multivariate dynamics and Synergetics.
We argue that although the former methodologies may 
be faster, it hides the cost of generating the data sets 
via the apriori training over several molecules. 
Our methodology, on the other hand, has no such
requirements. It is, in that sense, 
a standalone process, and can be executed from start to 
end in one go, without any prior training process. 

\section{Conclusion}
In this work, we demonstrate that the dynamics
of the CC nonlinear iterative scheme is dictated by a 
few significant excitations, whereas thousands of
cluster amplitudes are enslaved. Exploiting the basic 
principle of Synergetics, we employed a supervised ML 
scheme based on KRR model to express the enslaved
dependent amplitudes in terms of the independent master 
amplitudes. This leads us to develop a hybrid CC-ML 
algorithm where only the significant 
master amplitudes are selectively determined via 
BCH multi-commutator expansion. This leads to a
significant reduction in scaling for these iterations,
leading to enormous savings in the overall time taken 
for the computation. The enslaved amplitudes, on the 
other hand, are predicted via the KRR model as unique
functions of the master amplitudes. Our pilot 
numerical study demonstrates a sub $\mu E_h$ accuracy
of the predicted energy with 40\%-50\% reduction in 
overall computation time. The method is based on
physically motivated approximate schemes and does not 
require \textit{any} apriori training on different
molecules. This makes the algorithm standalone, and 
free of any hidden
cost. Furthermore, the algorithm is based on the CC
time-series dynamics, and hence it can easily be
generalized to include triple, quadruple
or higher excitations. Also, convergence can easily 
be further accelerated by using DIIS. One may further test
the efficacy of other supervised or unsupervised 
ML schemes as there are many readily available
plug-and-play ML models. This opens up the possibility of
interfacing different ML models in our algorithm, which 
will occupy us in near future.

\begin{acknowledgments}
RM acknowledges IIT Bombay Seed Grant, and Science and 
Engineering Research Board (SERB), Government of India, 
for financial support. 
\end{acknowledgments}
\section*{Data Availablity}
The data to support the findings of this study 
are available from the corresponding author upon
reasonable request.
%\section*{Conflict of interest}
%\bibliography{literature}% Produces the bibliography via BibTeX.

%merlin.mbs aipnum4-1.bst 2010-07-25 4.21a (PWD, AO, DPC) hacked
%Control: key (0)
%Control: author (8) initials jnrlst
%Control: editor formatted (1) identically to author
%Control: production of article title (0) allowed
%Control: page (1) range
%Control: year (1) truncated
%Control: production of eprint (0) enabled
%

\end{document}